\documentclass[%
superscriptaddress,
 preprint,
showpacs,preprintnumbers,
amsmath,amssymb,
aps,
prb,
]{revtex4-2}

\usepackage{graphicx}% Include figure files
\usepackage{dcolumn}% Align table columns on decimal point
\usepackage{bm}% bold math
\usepackage{amsmath}
\usepackage{upgreek}
\usepackage{hyperref}% add hypertext capabilities
\begin{document}

\title{Coexistence and manipulation of multiple singularities in a reconfigurable non-Hermitian metasurface}

\author{Xintong Shi}
\affiliation{School of Information Engineering, Nanchang University, Nanchang 330031, China}
\affiliation{School of Advanced Manufacturing, Nanchang University, Nanchang 330031, China}
\affiliation{College of Physical Science and Technology, Central China Normal University, Wuhan 430079, China}

\author{Yanjie Wu}
\affiliation{College of Physical Science and Technology, Central China Normal University, Wuhan 430079, China}
\affiliation{School of Electronic Science and Engineering, University of Electronic Science and Technology of China, Chengdu 611731, China}

\author{Rui Zhou}
\affiliation{College of Physical Science and Technology, Central China Normal University, Wuhan 430079, China}

\author{Zuxing Lu}
\affiliation{College of Physical Science and Technology, Central China Normal University, Wuhan 430079, China}

\author{Tengyu Li}
\affiliation{College of Physical Science and Technology, Central China Normal University, Wuhan 430079, China}

\author{Tingting Liu}
\affiliation{School of Information Engineering, Nanchang University, Nanchang 330031, China}
\affiliation{Institute for Advanced Study, Nanchang University, Nanchang 330031, China}

\author{Hai Lin}
\email{linhai@mail.ccnu.edu.cn}
\affiliation{College of Physical Science and Technology, Central China Normal University, Wuhan 430079, China}

\author{Qiegen Liu}
\email{liuqiegen@ncu.edu.cn}
\affiliation{School of Information Engineering, Nanchang University, Nanchang 330031, China}

\author{Shuyuan Xiao}
\email{syxiao@ncu.edu.cn}
\affiliation{School of Information Engineering, Nanchang University, Nanchang 330031, China}
\affiliation{Institute for Advanced Study, Nanchang University, Nanchang 330031, China}

\begin{abstract}
Non-Hermitian frameworks extend conventional Hermitian physics, offering a powerful paradigm for describing open systems. Central to this field are various singularities within the complex parameter space, such as exceptional points (EPs) and scattering zeros, which dictate exotic physical behaviors. As research shifts from isolated singularities toward multi-singularity interactions, conventional planar metasurfaces remain constrained by limited tuning dimensions. Here, we propose a mirror-coupled design that maps a metasurface into a quasi-high-dimensional parameter space. By employing a metallic plane to generate image resonators, this scheme multiplies the system degrees of freedom without increasing the number of physical resonators. Its implementation on a reconfigurable platform integrated with PIN diodes yields the coexistence and manipulation of an EP and multiple reflection zeros. Through simulations and microwave experiments, we characterize the dynamic evolution of these singularities and exploit their synergistic effects for two distinct applications. First, for tunable absorption, multiple reflection zeros are spectrally coordinated to achieve a near-perfect absorption band exceeding $99.9\%$ across the X-band, thereby dynamically suppressing target scattering. Second, for enhanced sensing, a reflection zero couples with the EP to form a hybrid singularity. This hybrid state inherits the power-law sensitivity of the EP while substantially boosting robustness against fluctuations, resolving the conventional trade-off between sensitivity and stability and simplifying detection to direct peak tracking rather than complex multimode eigenvalue fitting. Our work provides a general methodology to circumvent parameter competition among non-Hermitian singularities, opening new avenues for multifunctional metadevices across the electromagnetic spectrum.
\end{abstract}

\maketitle

%============================================================
\section{Introduction}
%============================================================
In mathematics, a singularity denotes the divergence of a function, whereas in physics, it signifies a critical state where system properties vanish or become indeterminate. Over the past decade, ushered in by the exploration of parity–time ($\mathcal{PT}$) symmetric non-Hermitian frameworks~\cite{ref1-Bender1998PRL,ref2-Bender1999JMP,ref3-Bender2007RPP}, a distinct class of non-Hermitian singularities, exceptional points (EPs), where eigenvalues and eigenstates coalesce, has ignited profound research interest across diverse wave systems~\cite{ref4-Assawaworrarit2017Nature,ref5-Choi2018NC,ref6-Miri2019Science,ref7-Ozdemir2019NatMat,ref8-Tang2020Science,ref9-Kononchuk2022Nature,Zheng2025Exceptional,Rong2026Arbitrary,Qin2026NatureRevMat}. Early endeavors predominantly exploited the localized physical attributes of isolated EPs for target-specific applications. These include leveraging the power-law scaling of eigenvalue splitting for enhanced trace sensing~\cite{ref10-Wiersig2016PRA,ref11-Chen2017Nature,ref12-Hodaei2017Nature,ref13-Xu2024NatNanotechnol,ref14-Mao2024SciAdv,ref15-NagChowdhury2025LPR,Bai_Wang_Zhang_Cui_2024}, utilizing branch-point characteristics for wavefront engineering~\cite{ref16-Song2021Science,ref17-Yang2024NC}, and harnessing parametric topological features for optical field manipulation~\cite{ref18-Schumer2022Science,ref19-Ergoktas2022Science}. As the field matures, the non-Hermitian landscape has expanded from closed-form eigenstate singularities to a broader family of spectral and scattering singularities. Prominent examples include bound states in the continuum (BICs), which manifest as radiationless singularities in the complex frequency plane~\cite{ref22-Sadreev2021RPP,Huang2023PhysRep}, and reflectionless or transmissionless scattering modes (RSMs and TSMs) that represent the fundamental zeros of the scattering matrix~\cite{Berkhout2019PerfectAbsorption,ref31-Zhang2025SciAdv,ref32-Shaibe2025PRR,Liu2025Bound}. These disparate singularities do not exist in isolation but rather form an interconnected topological ecosystem. Recent frontier studies have elucidated the dynamic evolution between BICs and EPs~\cite{ref23-Niu2024PRB,ref24-Wang2025LSA}, demonstrating that the merging of multiple BICs can spawn an ``exceptional BIC''\cite{ref25-CanosValero2025PRL}. Similarly, the coalescence of multiple scattering zeros in the complex plane can give rise to an EP~\cite{ref33-Sweeney2019PRL,ref34-Sweeney2020PRA}. Moreover, recent experiments on magnonic platforms have confirmed that scattering zeros and EPs share a unified topological classification framework characterized by intricate topological braiding~\cite{ref35-Rao2024NatPhys}. These breakthroughs signify a major paradigm shift in non-Hermitian physics, transitioning from the deterministic control of a single isolated singularity toward the interplay and manipulation of multiple coexisting singularities.

Metasurfaces, as artificial two-dimensional (2D) structures, possess exceptional electromagnetic wave manipulation capabilities unavailable in conventional devices. They have flourished across diverse fields including communications, beam steering, sensing, and imaging~\cite{Liu_Ma_Shao_Zhang_Yan_Ma_Zhang_Cui_2022,ref36-Wang2024TAP,ref37-Wu2025TMTT,ref38-Rosas2025Optica,ref39-Sun2026NatPhoton,ref40-Zhou2026ACSPhoton}, while simultaneously serving as an ideal platform for exploring non-Hermitian singularities~\cite{Lawrence_Xu_Zhang_Cong_Han_Zhang_Zhang_2014,Wang_Li_Zhang_Caihong_2017,Park_Lee_Baek_Ha_Lee_Min_Zhang_Lawrence_Kim_2020,ref51-Shi2025LPR}. However, realizing non-Hermitian singularities in metasurface platforms presents significant challenges. First, because system responses diverge at the singularity and vary substantially in its vicinity, precisely locating the exact singular state remains an intrinsic challenge~\cite{ref41-Heiss2012JPA,ref42-Shou2025npjNanophoton}. Second, this divergent nature results in an extremely low tolerance for fabrication errors, as unavoidable dimensional deviations and material parameter shifts can easily cause the system to deviate from the ideal singularity condition~\cite{ref42-Shou2025npjNanophoton,ref43-Li2023NatNanotechnol}. To address these issues, the integration of tunable elements to metasurfaces has emerged as an effective solution~\cite{Xiao2020JPhysD}. For instance, electrically controlled graphene gates enable the continuous tuning of chiral EPs~\cite{ref44-Baek2023LSA}. Similarly, optical pumping can modify the conductivity of semiconductors, such as silicon and germanium, to control the ohmic loss of resonators, thereby enabling ultrafast EP tuning, chirality inversion, and optical switching~\cite{ref45-He2023AdvSci,ref46-Masharin2024ACSNano,ref20-Yu2024AdvSci,ref21-He2025PRL}. While these approaches have proven highly effective for controlling a single isolated singularity, metasurfaces face a third, more fundamental limitation. Specifically, because the parameters governing different singularities are often independent or even mutually competitive~\cite{ref23-Niu2024PRB,ref25-CanosValero2025PRL,Colom2023LPR}, the synergistic manipulation of multiple singularities within a single metasurface encounters stringent physical constraints~\cite{ref23-Niu2024PRB,ref30-Wang2021Science}. Compared with bulk devices, metasurfaces as 2D structures offer fewer independent tuning dimensions, making it exceptionally difficult to simultaneously and independently control multiple system parameters such as resonant frequency, loss, and coupling strength. Because of these constraints, the integration of multiple non-Hermitian singularities within a single metasurface and the exploitation of their synergistic effects to enhance device performance have remained, to the best of our knowledge, largely unrealized.

In this work, we overcome these constraints by introducing a mirror-coupled metasurface design that operates within a quasi-high-dimensional parameter space. By placing a metallic plane beneath a pair of physical resonators, we leverage electromagnetic imagery to induce image resonators, thereby multiplying the available degrees of freedom without requiring additional physical resonators. This architectural expansion effectively lifts the intrinsic parameter competition that typically restricts multi-singularity systems. To further insulate the system from fabrication error-induced shifts, we implement this design on an active platform integrated with PIN diodes. This approach allows for post-fabrication electrical reconfiguration, enabling an EP and multiple reflection zeros to seamlessly coexist and evolve within a single device. Using full-wave simulations and microwave experiments in the X-band, we systematically characterize the interplay of these coexisting singularities across two distinct applications. For tunable absorption, the device can be dynamically toggled from a total-reflection state to a broadband, near-perfect absorption regime exceeding 99.9\%, where multiple reflection zeros are spectrally coordinated to suppress target scattering. For enhanced sensing, we address the notorious detection instability typically caused by EP divergence by coupling a reflection zero with the EP to form a hybrid singular state. This hybrid configuration inherits the power-law high sensitivity of the EP while drastically boosting measurement robustness against environmental fluctuations. Consequently, this strategy resolves the conventional trade-off between sensitivity and stability, streamlining the sensing readout into direct peak tracking instead of relying on complex multimode eigenvalue fitting. Our framework establishes a general methodology to expand effective parameters in lower-dimensional systems, paving the way for multifunctional non-Hermitian metadevices across the electromagnetic spectrum.

% ============================================================
\section{Results}
% ============================================================

% --- 1.1 Mirror-Coupled Model ---
\subsection{Mirror-coupled model for the non-Hermitian metasurface}
\label{sec:mirror-coupled}

\begin{figure}[htbp]
	\centering
	\includegraphics[width=0.57\textwidth]{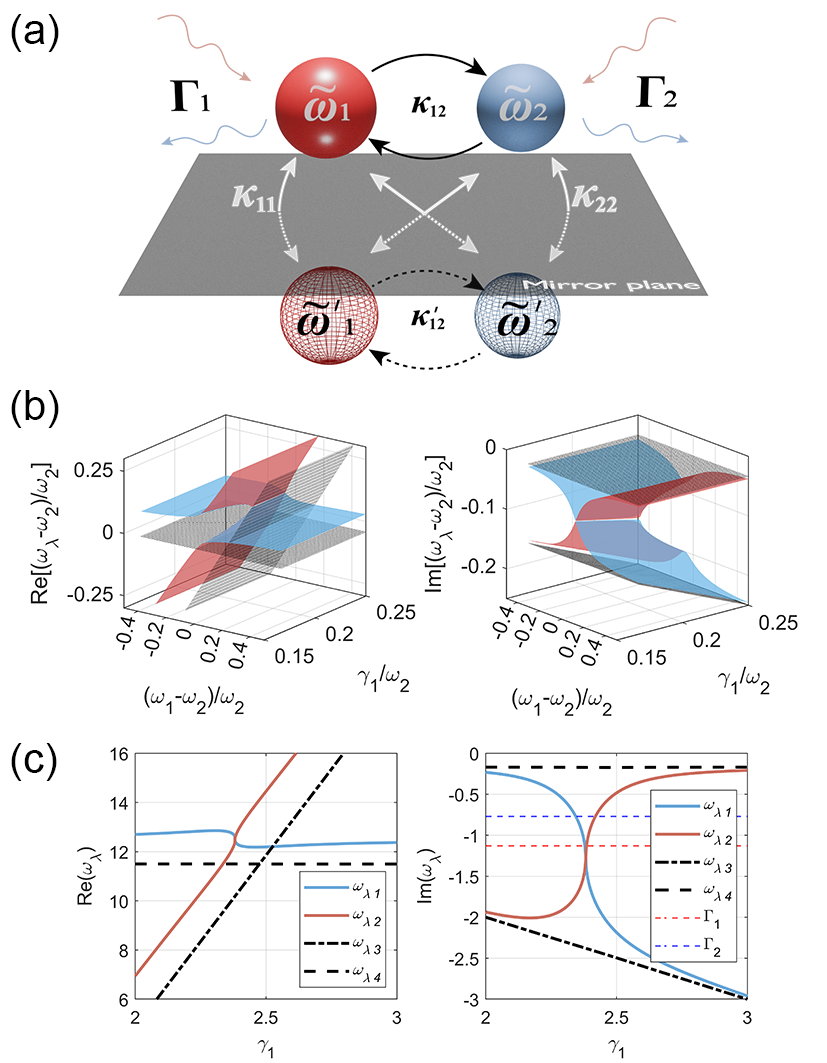}
	\caption{Conceptual schematic of the mirror-coupled non-Hermitian system and its eigenvalue evolution. 
		(a) The physical-image coupled framework consisting of physical resonators and their induced image resonators. 
		(b) The eigenvalue parameter space of the coupled system.  
		(c) The trajectories of eigenvalue evolution under parametric tuning of $\tilde{\omega}$, where the blue, red, and black solid curves map onto the respective surfaces, and the dashed lines denote the radiation losses of the resonators.}
	\label{fig:mirror_model}
\end{figure}

We begin by constructing a non-Hermitian coupled system based on a mirror-coupled scheme. The system consists of a pair of coupled physical resonators and a metallic plane. Specifically, one resonator (highlighted in red) features a tunable resonant frequency and loss, whereas the other (highlighted in blue) possesses a fixed resonant frequency and loss. Their complex frequencies are denoted as $\tilde{\omega}_1 = \omega_1 - i\gamma_1$ and $\tilde{\omega}_2 = \omega_2 - i\gamma_2$, respectively, where $\gamma_{1,2}$ represents the total loss of each resonator, comprising the radiation loss $\Gamma_{1,2}$ and the ohmic (intrinsic) loss $\gamma_{1,2} - \Gamma_{1,2}$. The coupling strength between the two physical resonators is denoted by $\kappa_{12}$. Upon introducing the metallic plane, according to the mirror principle, a pair of image resonators $\tilde{\omega}_1' = \omega_1' - i\gamma_1'$ and $\tilde{\omega}_2' = \omega_2' - i\gamma_2'$ are induced on the opposite side of the mirror. Each image resonator shares the identical complex frequency with its physical counterpart, i.e., $\tilde{\omega}_1' = \tilde{\omega}_1$ and $\tilde{\omega}_2' = \tilde{\omega}_2$. The mirror coupling strength between the two image resonators is denoted by $\kappa_{12}'$, and the coupling strengths between each image resonator and its corresponding physical resonator are denoted by $\kappa_{11}$ and $\kappa_{22}$, respectively. If the spacer is lossfree, the $\kappa_{12}' = \kappa_{12}$. This physical model is schematically illustrated in Fig.~\ref{fig:mirror_model}(a).

The effective Hamiltonian $\hat{H}$ of the mirror-coupled system is represented by a $4 \times 4$ matrix:
\begin{equation}
	\hat{H} = \sum_{i \in \{1,1',2,2'\}} \tilde{\omega}_i |i\rangle\langle i| + \sum_{i \neq j} \kappa_{ij} |i\rangle\langle j|.
	\label{eq:hamiltonian}
\end{equation}
Solving for the eigenvalues of this Hamiltonian yields two pairs of eigenvalues:
\begin{equation}
	\omega_{\lambda_{1,2}} = \frac{\tilde{\omega}_1 + \tilde{\omega}_2 + \kappa_{11} + \kappa_{22}}{2} \pm \frac{\sqrt{\mathrm{X}}}{2},
	\label{eq:eigenvalue34}
\end{equation}
where $\mathrm{X} = \kappa_{11}^2 - 2\kappa_{11}\kappa_{22} - 2\kappa_{11}\tilde{\omega}_1 - 2\kappa_{11}\tilde{\omega}_2 + 4\kappa_{12}^2 + 8\kappa_{12}\kappa_{12}' + \kappa_{22}^2 - 2\kappa_{22}\tilde{\omega}_1 + 2\kappa_{22}\tilde{\omega}_2 + 4\kappa_{12}'^{\,2} + \tilde{\omega}_1^2 + 2\tilde{\omega}_1\tilde{\omega}_2 + \tilde{\omega}_2^2$, and
\begin{equation}
	\omega_{\lambda_{3,4}} = \frac{\tilde{\omega}_1 + \tilde{\omega}_2 - \kappa_{22} - \kappa_{11}}{2} \pm \frac{\sqrt{\mathrm{Y}}}{2},
	\label{eq:eigenvalue12}
\end{equation}
where $\mathrm{Y} = \kappa_{11}^2 - 2\kappa_{11}\kappa_{22} - 2\kappa_{11}\tilde{\omega}_1 + 2\kappa_{11}\tilde{\omega}_2 + 4\kappa_{12}^2 - 8\kappa_{12}\kappa_{12}' + \kappa_{22}^2 + 2\kappa_{22}\tilde{\omega}_1 - 2\kappa_{22}\tilde{\omega}_2 + 4\kappa_{12}'^{\,2} + \tilde{\omega}_1^2 + 2\tilde{\omega}_1\tilde{\omega}_2 + \tilde{\omega}_2^2$.
Eqs.~(\ref{eq:eigenvalue34}) and (\ref{eq:eigenvalue12}) can be rewritten in a more familiar form:
\begin{equation}
	\omega_{\lambda_{1,2}} = \frac{\tilde{\omega}_1 + \tilde{\omega}_2 + \kappa_{11} + \kappa_{22}}{2} \pm \frac{\sqrt{(\kappa_{11} - \kappa_{12} + \tilde{\omega}_1 - \tilde{\omega}_2)^2 + 4(\kappa_{12} + \kappa_{12}')^2}}{2},
	\label{eq:eigenvalue34_simple}
\end{equation}
\begin{equation}
	\omega_{\lambda_{3,4}} = \frac{\tilde{\omega}_1 + \tilde{\omega}_2 - \kappa_{22} - \kappa_{11}}{2} \pm \frac{\sqrt{(\kappa_{11} - \kappa_{12} - \tilde{\omega}_1 + \tilde{\omega}_2)^2 + 4(\kappa_{12} - \kappa_{12}')^2}}{2}.
	\label{eq:eigenvalue12_simple}
\end{equation}

Fig.~\ref{fig:mirror_model}(b) depicts the self-intersecting Riemann surfaces formed by these two pairs of normalized eigenvalues, which are obtained by sweeping the resonant frequency and damping rate of one resonator while keeping the coupling strengths fixed. The left and right panels display the real and imaginary parts of the eigenvalues, respectively. The red and blue surfaces correspond to the first pair of eigenvalues and support an EP, whereas the gray surfaces correspond to the second pair and exhibit no coalescence. Eqs.~(\ref{eq:eigenvalue34_simple}) and (\ref{eq:eigenvalue12_simple}) demonstrate that the eigenvalue degeneracy does not correspond to a standard second-order EP. Although this EP is formed from a single pair of eigenvalues, a first-order residual term remains in the eigenvalue splitting expansion $\sqrt{\mathrm{X}}$. 

Figure~\ref{fig:mirror_model}(c) plots the evolution trajectories of the coupled-system's eigenvalues with respect to the tunable resonator. Here, the total loss $\gamma_1$ of the tunable resonator serves as the independent variable, which is dynamically tailored via our parallel structural design (see Supplementary Material 1 for technical details). The left and right panels show the real and imaginary parts of the eigenvalues, respectively. As $\gamma_1$ increases, $\mathrm{Re}(\omega_{\lambda_2})$ and $\mathrm{Re}(\omega_{\lambda_3})$ increase rapidly before genuine crossing, while $\mathrm{Re}(\omega_{\lambda_1})$ and $\mathrm{Re}(\omega_{\lambda_4})$ remain relatively steady. By optimizing the coupling parameters, the system can be engineered to a point where the real and imaginary parts of $\omega_{\lambda_1}$ and $\omega_{\lambda_2}$ simultaneously degenerate, thereby establishing an EP. At this critical point, the system satisfies $(\kappa_{11} - \kappa_{12} + \tilde{\omega}_1 - \tilde{\omega}_2)^2 + 4(\kappa_{12} + \kappa_{12}')^2 = 0$.

On the other hand, the total loss of each resonator can be divided into radiation loss and ohmic loss. The radiation loss is explicitly governed by the structural geometry, whereas the ohmic loss depends on the resistivity and dielectric loss of the constituent materials. In the right panel of Fig.~\ref{fig:mirror_model}(c), the red and blue dashed lines represent the radiation losses $\Gamma_{1,2}$ of the $\tilde{\omega}_1$ and $\tilde{\omega}_2$ resonators, respectively. When the ohmic loss is tuned to equal the radiation loss~\cite{Wang_Chen_Zhang_Zeng_Zhang_Liu_Shi_Zi_2019,ref49-Wang2020PRB}, the system satisfies the critical coupling condition for perfect absorption. For a reflective metasurface, this critical coupling manifests as multiple points with zero reflectance in the reflection spectrum. Although directly extracting the radiation loss from the spectrum is challenging, particularly in multimode coupled systems, it can be elegantly deduced from the critical coupling condition~\cite{ref33-Sweeney2019PRL,ref34-Sweeney2020PRA,Liu2023PRB}. For a single-port system, the complex frequency of a reflection zero reduces to a purely real value at critical coupling, i.e., $\gamma = 2\Gamma$. Through structural optimization, this condition can be achieved, allowing both reflection zeros and an EP to coexist within a single metasurface platform. The interplay among multiple coexisting singularities significantly expands the functional diversity of non-Hermitian metadevices.

% --- 1.2 Characterization of Singularities ---
\subsection{Characterization of singularities in the non-Hermitian metasurface}
\label{sec:characterization}

\begin{figure}[htbp]
	\centering
	\includegraphics[width=0.85\textwidth]{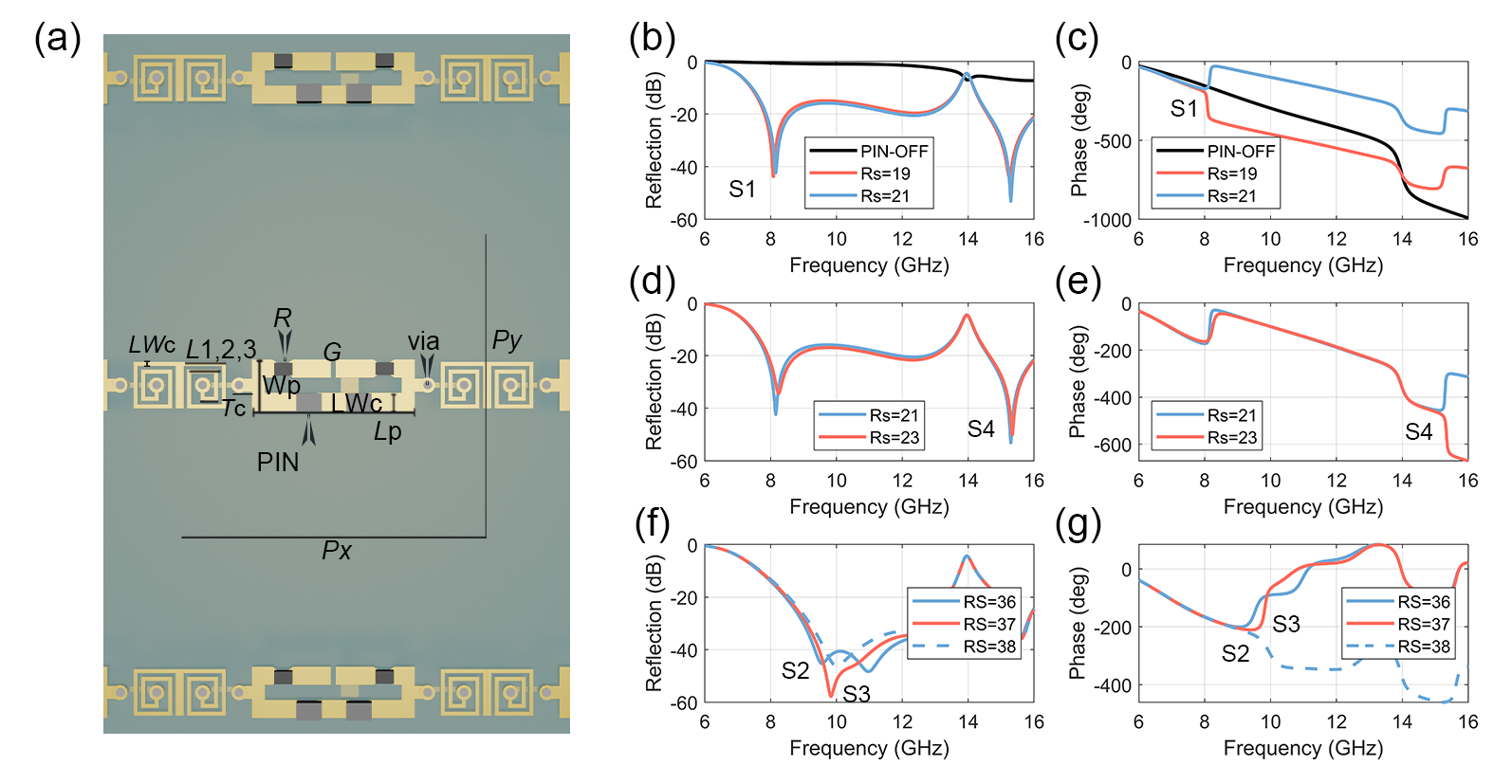}
	\caption{Full-wave electromagnetic simulations of the coexisting multiple singularities in the non-Hermitian metasurface. 
		(a) The geometrical configuration of the optimized metasurface unit cell. The structural dimensions include a periodicity of $P_x = P_y = 11\text{ mm}$, a parallel structure length of $L_\mathrm{p} = 5.35\text{ mm}$ with a width of $W_\mathrm{p} = 2\text{ mm}$, a linewidth of $W_\mathrm{C} = 0.7\text{ mm}$, a split gap width of $G = 0.2\text{ mm}$, and a tip length of $T_\mathrm{c} = 0.5\text{ mm}$. The lengths of the nested spiral arms from outer to inner are defined as $L_{1,2,3} = 1.8, 1.2, 0.68\text{ mm}$ with a linewidth of $LW_s = 0.15\text{ mm}$, while the mirror spacing is fixed at $11\text{ mm}$. The integrated active components comprise an SMP1345-040LF PIN diode and a $120\ \Omega$ biasing resistor, both patterned on a Rogers RO4350B substrate with a relative permittivity of $\varepsilon_r = 3.48 \pm 0.05$ and a loss tangent of $\tan\delta = 0.0037$ at $10\text{ GHz}$. 
		The simulated reflectance amplitude and corresponding phase spectra of (b, c) the singularity S1, (d, e) the singularity S4, and (f, g) the coexisting singularities S2 and S3.}
	\label{fig:sim_char}
\end{figure}

We utilize finite-element simulations to optimize the metasurface geometric parameters and perform microwave experiments to characterize the designed singularities. The metasurface is constructed by connecting two distinct shaped structures in series to form a pair of physical resonators. A metallic backplane serves as the mirror plane, establishing a physical-image coupled system under unit cell boundary conditions. The unit-cell geometry of the metasurface is illustrated in Fig.~\ref{fig:sim_char}(a). The corresponding simulation results are presented in Figs.~\ref{fig:sim_char}(b)--\ref{fig:sim_char}(g), where the electromagnetic response is dynamically modified by varying the resistance of the integrated PIN diodes. Under TM-polarized Floquet-mode excitation, a parametric sweep over the PIN diode resistance is performed to map the complete reflection spectra within the target parameter space. 

To elucidate the distribution characteristics of these non-Hermitian singularities, we extract and plot the reflection amplitudes and phases at several key resistance values. Four distinct singular states, comprising three reflection zeros and one EP, are identified and labeled as S1 to S4 in order of increasing frequency. When the PIN diode is in the OFF state, the metasurface operates in a total-reflection regime, as indicated by the black curves in Figs.~\ref{fig:sim_char}(b) and \ref{fig:sim_char}(c), maintaining a high reflectance above $-1$\,dB across the entire X-band. At a diode resistance of $R_\mathrm{S} = 21\,\Omega$, the system approaches the first singularity, S1, with the corresponding profiles capturing the reflectance behavior on both sides of the zero point ($R_\mathrm{S} = 19\,\Omega$ and $R_\mathrm{S} = 21\,\Omega$). Singularity S4 emerges within a similar resistance range but at a higher frequency, localized between $R_\mathrm{S} = 21\,\Omega$ and $R_\mathrm{S} = 23\,\Omega$ as shown in Figs.~\ref{fig:sim_char}(d) and \ref{fig:sim_char}(e). Driven by its proximity to the EP, singularity S2 exhibits extreme sensitivity to parametric variations, manifesting a pronounced phase singularity near 9.5\,GHz. Singularity S3 corresponds to the exact EP, with its local spectra evaluated at $R_\mathrm{S} = 36\,\Omega$ and $R_\mathrm{S} = 38\,\Omega$. In this immediate vicinity, the metasurface undergoes a phase transition from the $\mathcal{PT}$-symmetric phase to the $\mathcal{PT}$-broken phase, thereby confirming the presence of the EP.

\begin{figure}[htbp]
	\centering
	\includegraphics[width=0.57\textwidth]{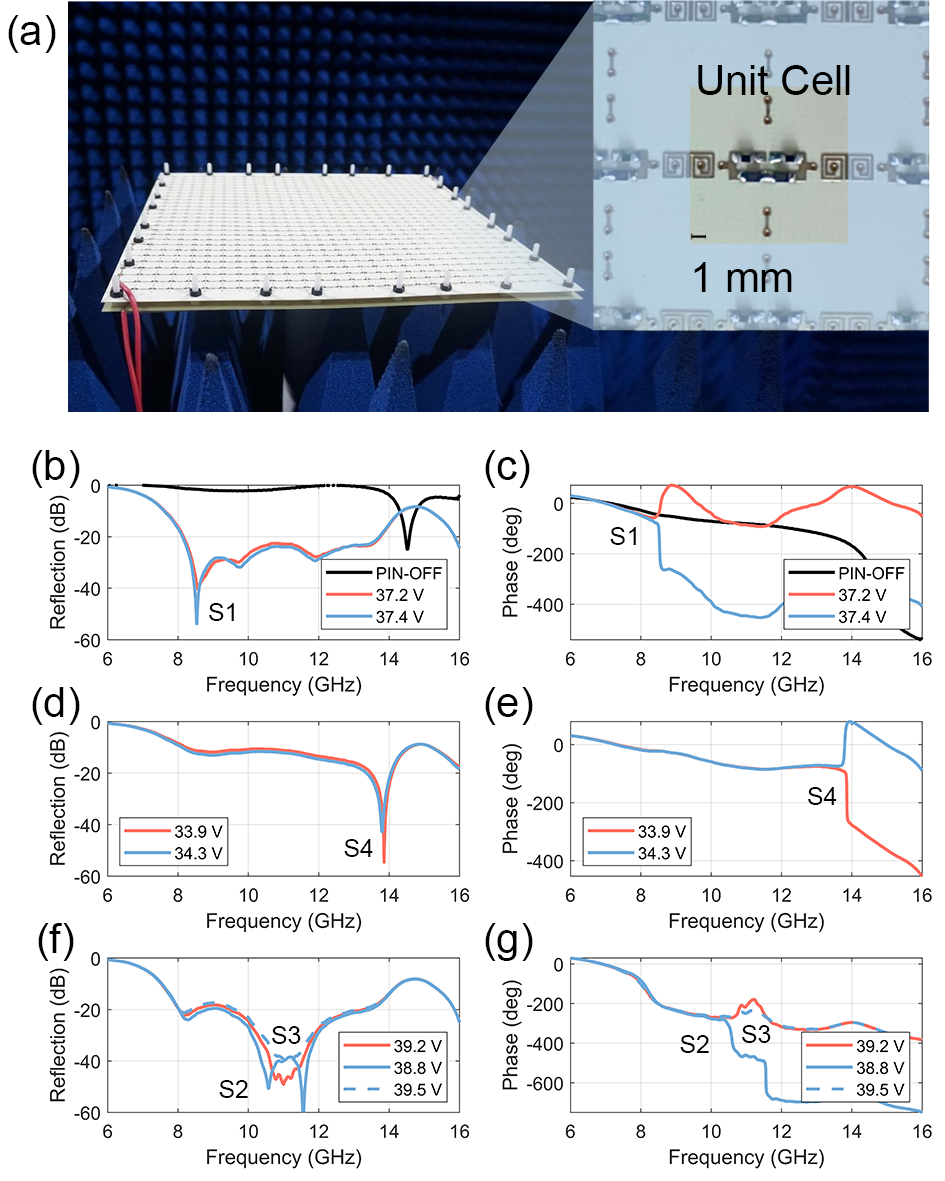}
	\caption{Experimental characterization of the coexisting multiple singularities in the non-Hermitian metasurface. 
		(a) The photograph of the fabricated metasurface sample on the arch-field measurement stage, where the right panel shows a magnified view of the unit cell. 
		The measured reflectance amplitude and corresponding phase spectra of (b, c) the singularity S1, (d, e) the singularity S4, and (f, g) the coexisting singularities S2 and S3.}
	\label{fig:exp_char}
\end{figure}

The metasurface sample is fabricated using standard printed circuit board (PCB) technology, comprising a functional structural layer and a reflective layer. The structural layer employs a Rogers RO4350B substrate with a thickness of 0.508\,mm. The metallic patterns and metallized vias are realized via chemical etching and immersion gold plating, whereas the resistors and PIN diodes are mounted using automatic reflow soldering. The reflective layer consists of a continuous, fully copper-clad plane without any patterning. The overall metasurface dimensions are $300 \times 300$\,mm$^{2}$, containing an array of $25 \times 25$ unit cells. The structural and reflective layers are connected and fixed using polymer screws, where the air spacer thickness is precisely controlled by inserting precision shims to tune the interlayer distance. The fabricated metasurface is shown in Fig.~\ref{fig:exp_char}(a), with the right inset showing a magnified view of the local lattice array where the unshaded region delineates a single unit cell.

During the experiments, a digital direct current (DC) power supply is employed to control the bias voltage applied to the metasurface, thereby modulating the underlying resistance of the PIN diodes (see Supplementary Material 2 for technical details) to achieve dynamic singularity reconfiguration. By synchronizing the DC power supply with a Cyear 3671G vector network analyzer (VNA), the reflection spectra are continuously recorded across the entire preset voltage range. According to the manufacturer datasheet, the turn-on voltage of an individual diode is 0.68\,V, implying that the threshold voltage for the entire array is approximately 34\,V, with full saturation reached at 55\,V. Through systematic experimental calibration, the bias voltage sweep is optimized within the range of 33--40\,V, which successfully maps out all targeted singularities within the available parameter space.

Figures~\ref{fig:exp_char}(b)--\ref{fig:exp_char}(g) illustrate the measured reflection spectra in the vicinity of the singularities. When the applied bias voltage is set to 37.2\,V, the system closely approaches singularity S1, where the reflectance drops drastically and the phase undergoes an abrupt transition. At a bias voltage of 33.9\,V, the system reaches the neighborhood of singularity S4, capturing a similarly steep phase jump. At 38.8\,V, the response tracks close to singularity S2, again exhibiting a distinct phase jump. Finally, at 39.2\,V, the system is tuned closest to the EP (S3). On either side of this bias voltage, the system transitions between the $\mathcal{PT}$-symmetric phase and the $\mathcal{PT}$-broken phase, in strong agreement with the numerical simulations.

As anticipated, practical imperfections such as fabrication errors, soldering-induced perturbations, and non-ohmic contacts inevitably influence the experimental metasurface performance, leading to minor shifts in the singularity positions compared to the numerical predictions. These deviations reflect the inherent sensitivity of non-Hermitian spectral states to structural variations. Nevertheless, our electrical reconfiguration strategy effectively compensates for these geometric and material deviations, as demonstrated by the clear experimental observation of all simulated singularities. It is worth noting, however, that this robust tolerance is conditional: the structural tolerances must remain within a manageable threshold that avoids severe physical discontinuities or unintended short circuits in the biasing grid. If the fabrication defects are sufficiently severe to alter the fundamental resonant modes of the entire structure, post-fabrication reconfiguration would eventually become ineffective.

\begin{figure}[htbp]
	\centering
	\includegraphics[width=0.85\textwidth]{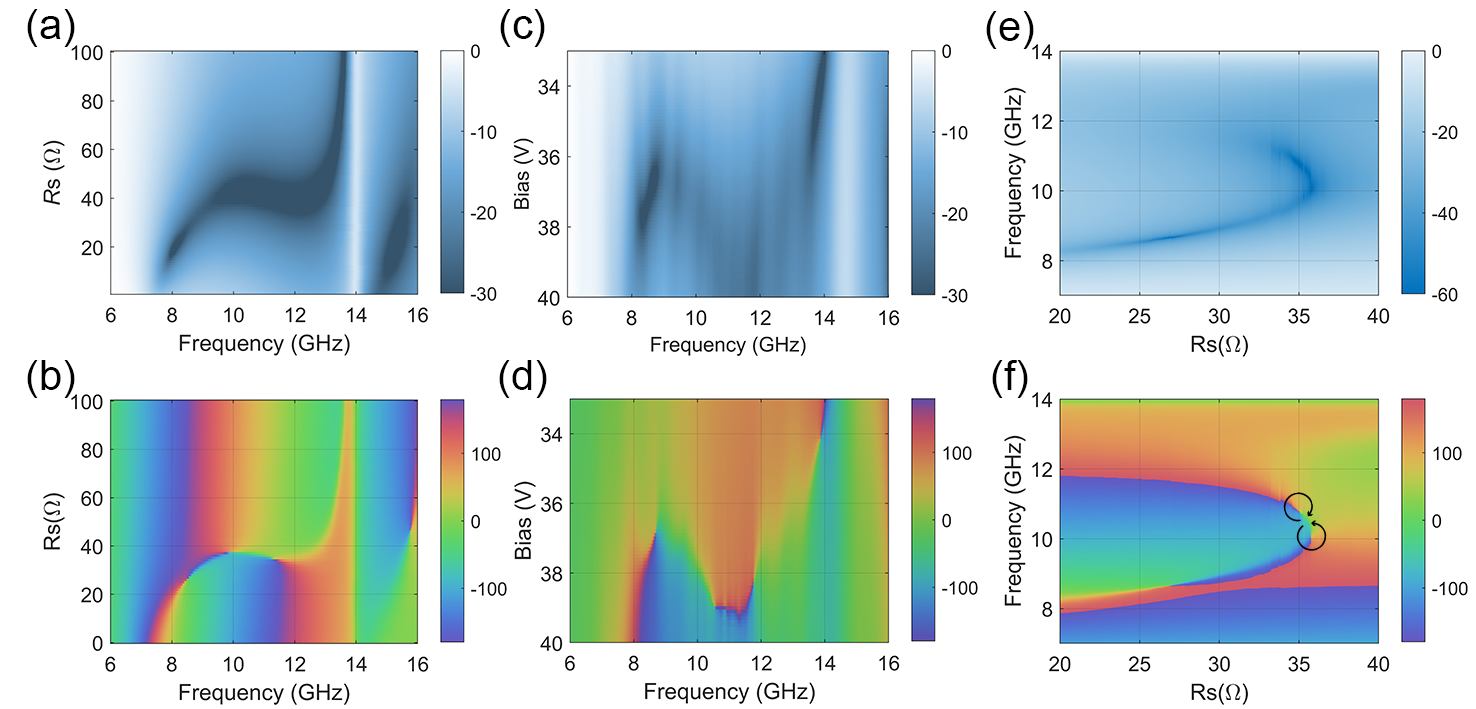}
	\caption{Comparison between simulated and experimentally measured reflectance profiles. The reflectance amplitude and corresponding phase spectra obtained via (a, b) the global full-wave simulations, (c, d) the experimental measurements, and (e, f) the targeted simulations in the immediate vicinity of the EP.}
	\label{fig:full_comp}
\end{figure}

A comprehensive comparison between the simulated and experimental reflectance is mapped out in Figs.~\ref{fig:full_comp}(a)--\ref{fig:full_comp}(d). In the simulations, the resistance is continuously swept from 0 to 100\,$\Omega$, while the experimental bias voltage spans 33 to 40\,V to establish a consistent parameter mapping, the details of which are provided in Supplementary Material 2. When the bias voltage remains below the diode turn-on threshold, the metasurface behaves as a near-perfect reflector across the entire X-band. As the bias voltage increases, the reflectance gradually drops, manifesting a broadband suppression profile centered at approximately 38 V, where the minimum reflectance drops below $-30$\,dB. Concurrently, the phase distribution reveals multiple topological phase-singularity regions localized near these reflectance minima, whose centers track the S1, S2, S3, and S4 states in order of increasing frequency. The experimental results demonstrate strong agreement with the simulations. Note that the reflectance map in Fig.~\ref{fig:full_comp}(a) excludes the fully biased OFF state, which can be verified via the black curve in Fig.~\ref{fig:sim_char}(b) and Fig.~S11 of the Supplementary Material 4.

Figures~\ref{fig:full_comp}(e) and \ref{fig:full_comp}(f) illustrate the reflection spectra and phase profiles when the spacer thickness is adjusted to 5.8\,mm, capturing a continuous evolution of the $\mathcal{PT}$ phase transition. The results clearly reveal the merging of two modes. Specifically, when the spectrum displays two distinct resonant dips, the system resides within the $\mathcal{PT}$-symmetric phase. As $R_\mathrm{S}$ increases, the system crosses the EP and enters the $\mathcal{PT}$-broken phase. Since $\Delta\gamma_{1,2'} > 2\kappa_{12}'$, the spectrum subsequently collapses into a single resonance peak, which is clearly captured by the line traces in Figs.~\ref{fig:sim_char} and \ref{fig:exp_char}. The phase characteristics further indicate that two reflection zeros undergo near-annihilation at this point, each possessing mutually opposite topological invariants~\cite{ref34-Sweeney2020PRA,ref19-Ergoktas2022Science,Liu2023PRB}. A comprehensive analysis regarding the singularity trajectories and their underlying topological invariants is provided in Supplementary Material 3.

% --- 1.3 ------application---
\subsection{Applications for tunable absorption and enhanced sensing}
\label{sec:application} 

Here, we demonstrate the practical applications of dynamically manipulating multiple coexisting singularities through two representative applications, namely tunable absorption and enhanced sensing. By spectrally coordinating two adjacent reflection zeros, their inherent reflectionless characteristics are harnessed to form a broadband and near-perfect absorption profile that outperforms conventional single-resonance designs. Alternatively, when a reflection zero is brought into close spectral proximity with an EP, their mutual coupling spawns a hybrid singular state. Compared with conventional single-singularity sensing protocols, this hybrid configuration inherits the power-law high sensitivity of the EP while substantially boosting the environmental robustness of the measurement.

We first explore the manipulation of multiple singularities for expanding tunable absorption bandwidths. As mapped out across the complete reflection profiles in Figs.~\ref{fig:full_comp}(a) and \ref{fig:full_comp}(c), the electromagnetic response exhibits a wide tuning range from the onset to full diode saturation. Below the turn-on threshold voltage, the metasurface resides in a total-reflection regime. As the bias voltage rises, the PIN diodes are progressively forward-biased and their resistance drops, causing the reflectance to gradually diminish before experiencing a partial recovery. Near 40\,$\Omega$, a broadband absorption band spanning approximately 5\,GHz emerges, completely covering the target X-band with an absorptivity exceeding 99.9\%. In contrast, at alternative resistance states, while deep absorption dips persist at isolated reflection zeros, the operational bandwidth remains heavily constrained. This clear distinction demonstrates that through the synergistic effects of multiple singularities, the metasurface can achieve dynamic switching between near-perfect reflection and broadband perfect absorption. To achieve a polarization-insensitive response, the unit-cell topology is rotated by 90$^\circ$ and mirrored with respect to the central plane of the physical resonator substrate, as detailed in Supplementary Fig.~S7 within Supplementary Material 2. The complete experimental data and dynamic reflection profiles for the other polarization direction are provided in Supplementary Material 4.

\begin{figure}[htbp]
	\centering
	\includegraphics[width=0.85\textwidth]{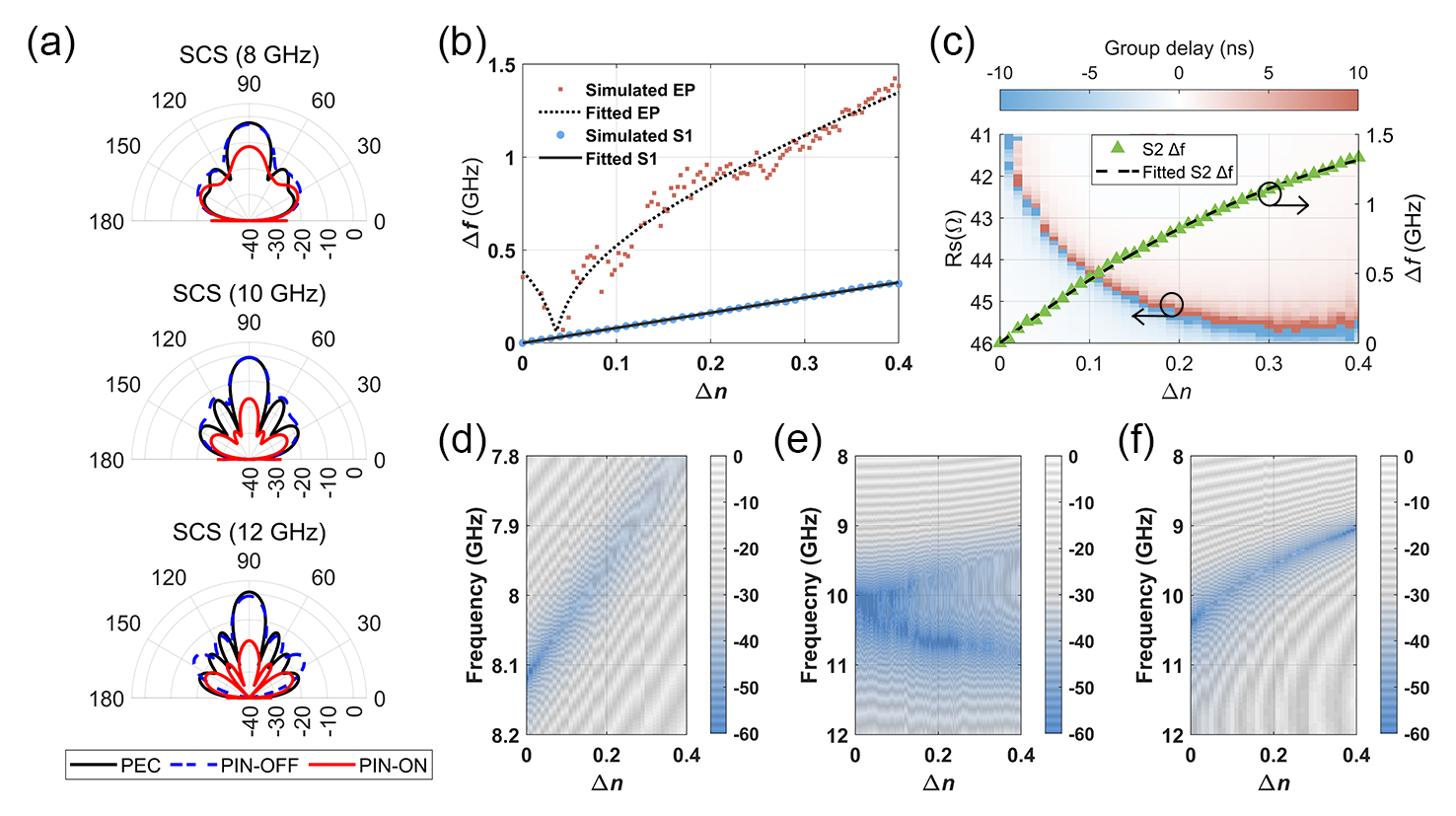}
	\caption{Application demonstrations of the multifunctional non-Hermitian metasurface. 
		(a) The scattering reduction profiles at $8$, $10$, and $12\text{ GHz}$ evaluated under PEC reference, PIN-OFF, and PIN-ON conditions. 
		The sensing parameter extractions under refractive index perturbations including (b) the fitting curves for the EP frequency splitting and the singularity S1 frequency shift, and (c) the frequency shift and resistance-tracking trajectories for the hybrid singularity S2.
		The perturbation-induced response spectra capturing (d) the singularity S1 frequency shift, (e) the EP frequency splitting, and (f) the hybrid singularity S2 frequency shift.}
	\label{fig:applications}
\end{figure}

A primary application scenario of such a broadband absorption lies in suppressing target scattering. To evaluate this performance, we analyze a finite-sized metasurface array under open boundary conditions and compare its scattering signature with that of a perfect electric conductor (PEC) plate of identical dimensions. Fig.~\ref{fig:applications}(a) depicts the simulated scattering patterns of an $8 \times 8$ unit-cell array across key sampled frequencies, comparing the PEC plate, the PIN-OFF state, and the PIN-ON state from top to bottom. In the PIN-OFF state, the scattering profile of the metasurface closely mirrors that of the bare PEC plate. Conversely, when the PIN diodes are fully turned on, the scattering intensity is suppressed by over 90\% across all sampled frequencies, reaching a reduction exceeding 99\% at both 10\,GHz and 12\,GHz. The angular stability of this scattering reduction is further quantified and discussed in Supplementary Material 5.

The second major application centers on enhanced trace sensing. To mimic a target analyte, a standard evaluation framework is adopted by introducing a dielectric slab atop the metasurface platform~\cite{ref38-Rosas2025Optica, ref50-Beruete2020AOM}. Specifically, a 1-mm-thick dielectric slab is inserted between the physical resonators and the metallic plane. The refractive index is varied from $n = 1.0$ to $1.4$ ($\Delta n = 0.4$), and the corresponding reflection profiles are extracted via parametric sweeps. Fig.~\ref{fig:applications}(b) presents the simulated data along with the theoretical fits for the frequency near the EP and the S1 reflection zero point, where the EP and S1 frequency responses are designated by red and blue scatters, respectively. The S1 frequency shift scales linearly with the refractive index variation, whereas the EP splitting follows a characteristic power-law dependence, the fitting details of which are provided in Supplementary Material 6. Crucially, over the identical refractive index span, the frequency splitting of the EP significantly exceeds the linear shift of S1, confirming its superior sensitivity to external perturbations. Figs.~\ref{fig:applications}(d) and \ref{fig:applications}(e) plot the evolution of the full reflection spectra in the vicinity of S1 and the EP, respectively. For the S1 state, an increasing analyte refractive index induces a uniform redshift of the resonant dip accompanied by a slight elevation in reflectance. In contrast, the EP spectrum experiences a rapid splitting into two distinct resonant dips as the perturbation intensifies.

Notably, the fluctuations of the S1 data points remain remarkably small, whereas the EP data points exhibit more pronounced instability. To further elucidate this feature, we plot the reflection spectrum using a gray-and-white striped colorbar, which effectively highlights subtle spectral variations. The sensitivity of the EP to underlying system parameters is clearly reflected in these maps: well away from the EP regime, specifically between 8 and 9\,GHz, the grayscale fringes remain highly continuous and stable, whereas they become disordered in the immediate vicinity of the EP. In contrast, the stripes near S1 remain highly stable. The physical origin of this fluctuation stems from mesh discretization artifacts introduced during the numerical sweep, which serve as a numerical analog to experimental fabrication errors. Although these are merely weak numerical fluctuations, they are sufficient to produce observable traces in the spectrum after amplification by the EP. Consequently, high sensitivity and measurement stability appear to be mutually exclusive in conventional EP systems, defining a fundamental trade-off that our framework explicitly addresses.

Figure~\ref{fig:applications}(c) plots the frequency shift of the hybrid singularity S2 as a function of the refractive index variation, with the green triangles and the black dashed line representing the simulated data and the theoretical power-law fit, respectively. As the refractive index scales from 0 to 0.4, the hybrid singularity exhibits a power-law frequency scaling that mirrors the high sensitivity of the conventional EP. Crucially, the data fluctuation is drastically minimized compared to the isolated EP, a stabilization further validated by the highly continuous and uniform grayscale fringes depicted in the full spectra of Fig.~\ref{fig:applications}(f). Taking $\Delta n = 0.2$ as a benchmark, the conventional EP splitting yields a sensitivity of 4.25\,GHz/RIU, whereas our hybrid singularity achieves a comparable sensitivity of 4.08\,GHz/RIU while substantially improving the detection stability. To evaluate practical viability, lake water samples with varying algae concentrations are prepared via binary dilution and tested, successfully validating the real-world applicability of this hybrid sensing scheme, with the detailed preparation protocols and sensing readouts compiled in Supplementary Material 8. Unlike conventional EP sensors that necessitate complex multimode eigenvalue fitting, our hybrid scheme relies entirely on direct, single-peak tracking. This approach is intrinsically faster, significantly more accurate, and fully compatible with standard commercial sensing architectures, as compared comprehensively in Supplementary Material 6.

Furthermore, adjusting the bias voltage applied to the metasurface to match the spectral singularities and dynamically track their positions can likewise serve as an effective characterization methodology for perturbations. Fig.~\ref{fig:applications}(c) illustrates the results of using PIN diode resistance to dynamically track the group delay profile of the phase singularity S2, where the color transitions delineate phase leads and lags. This group delay representation effectively compresses the data dimensionality, with the underlying raw profiles compiled in Supplementary Material 7. With the increase of the analyte refractive index, the required PIN diode resistance for singularity matching follows a well-defined nonlinear trajectory. These simulation results demonstrate that both the frequency and the PIN diode resistance can serve as independent, unique metrics to quantify changes in the analyte index. By mapping the resistance-to-voltage calibration provided in Supplementary Material 2, a direct analytical bridge is established between the bias voltage and the external perturbation parameters. 
Therefore, integrating distinct classes of non-Hermitian singularities within a reconfigurable metasurface circumvents the performance bottlenecks inherent to single-singularity protocols, while electrical tunability enables the metadevice to switch dynamically between optimized operational modes for diverse real-world scenarios.

% % ============================================================
\section{Discussion}
% ============================================================

In conclusion, we have developed a non-Hermitian framework based on a mirror-coupled scheme, successfully demonstrating the coexistence and manipulation of multiple singularities within a metasurface platform. By incorporating PIN diodes as active tuning elements, the system effectively compensates for post-fabrication shifts induced by geometric imperfections, while simultaneously enabling dynamic reconfiguration of the singularities. Building upon this platform, leveraging the synergistic effects of the coexisting multiple singularities allows for the realization of broadband tunable absorption and high-performance sensing. This highly promising scheme features a sophisticated physical mechanism: by effectively mapping the system onto a higher-dimensional parameter space without introducing additional physical resonators, this design successfully increases the effective dimension within a compact footprint, which eliminates the inherent parameter competition among distinct singular states and unlocks enhanced synergistic effects, thereby validating the capacity of this strategy to enrich multifunctional non-Hermitian devices. It is worth noting that the mirror-coupled model introduced here is inherently scalable beyond the microwave regime and independent of the specific geometric realization, which can be readily translated to electronic circuits, photonic crystals, and plasmonic and dielectric structures. Extending this concept to the terahertz or visible frequency bands requires only scaling the resonant geometries down to the appropriate micro- or nanoscale dimensions, combined with replacing the active diodes with phase-change or tunable materials such as liquid crystals, $\mathrm{Ge_2Sb_2Te_5}$, or doped semiconductors~\cite{ref27-Tian2019NC, ref52-Qiu2021Nanotechnology, ref53-He2024LSA, ref54-Zhang2025AOM, ref55-Fan2025LPR}.

\begin{acknowledgments}
	
This work was supported by the National Natural Science Foundation of China (Grants No. 62571212, No. 12364045, No. 12264028, and No. 12304420), the Natural Science Foundation of Jiangxi Province (Grants No. 20253BAC260002 and No. 20262BAC240253), the Fundamental Research Funds for the Central Universities (Grant No. JC2026TS-001), and the Young Elite Scientists Sponsorship Program by JXAST (Grant No. 2025QT04).

The authors thank Prof. Yangjie Liu at Hubei University for valuable discussions, and Weilong Wang and Yunfei Li for assistance with the experiments.
	
\end{acknowledgments}

%\bibliography{references}


\end{document}